%Paper: hep-ph/9303263
%From: Takahiro Kubota <KUBOTA%JPNYITP.BITNET@pucc.Princeton.EDU>
%Date: Tue, 16 Mar 93 20:36:35 JST

\magnification1200
\rightline{OS-GE     32-93}
\rightline{March 12, 1993}
\vskip3.5cm
\centerline{{\bf Lee-Quigg-Thacker Bounds for  Higgs
Boson Masses}}
\centerline{{\bf in a Two-Doublet Model}}
\vskip3.5cm
\centerline{Shinya KANEMURA, Takahiro KUBOTA and
Eiichi TAKASUGI}
\vskip1cm
\centerline{{\it
Institute of Physics, College of General Education }}
\centerline{{\it
Osaka University, Toyonaka, Osaka 560, Japan}}
\vskip2cm
\centerline{{\bf Abstract}}
Upper bounds for neutral as well as charged Higgs boson
masses in a two-doublet model are obtained on the basis
of tree unitarity conditions \`a la Lee, Quigg and
Thacker.  A wide variety of scattering processes are
considered so extensively that our bounds are more
restrictive than those obtained previously for neutral
Higgs bosons and are also of a new kind for charged Higgs
boson.  It is argued that at least one of the Higgs bosons
should be lighter than 580 ${\rm GeV}/c^{2}$.
\vfill\eject
\noindent
{\bf 1. Introduction}

Although the success of the standard $SU(2)_{W}\times U(1)_{Y}$
gauge theory  of the electroweak interactions is overwhelming,
the Higgs boson sector in charge of spontaneous symmetry
breaking has so far eluded experimental verification and is still
a mystery.  We all agree that the Higgs boson is one of
our central concerns of the present  experimental search and
will be so even more in the future colliders, JLC, SSC and LHC.

It has been known  by now rather well that the mass of the Higgs
boson, which is proportional to the Higgs quartic coupling,
may be bounded from above, provided that the quartic coupling
is not so large as to violate the validity of perturbative
calculations [1,2]. In fact in the minimal standard model
with a single Higgs doublet,  Lee, Quigg and Thacker (LQT) [1]
deduced the constraint from the  perturbative  unitarity which
turned out to be $m_{h}<(8\pi \sqrt{2}/3G_{F})^{1/2} \equiv M_{LQT}
\sim 1 {\rm TeV}/c^{2}$ , where $G_{F}$ is the Fermi constant and
$m_{h}$ is the mass of a neutral Higgs boson.

Extension of their type of analyses was considered by several
authors in the presence of more than one Higgs doublet.  There
are several motivations for  increasing the number of Higgs
bosons; supersymmetric extension of the standard model, a model
of spontaneous CP-violation due to Higgs sector [3], the
Peccei-Quinn mechanism [4] and so forth.    There is also
a surge of phenomenological interest in the two
 Higgs doublet model in  recent literatures [5].

Casalbuoni et al. [6] have raised a question which has close
bearings on  LQT's, namely,
at what energy strong interaction phenomena would start to show
 up whenever one or more Higgs masses are sufficiently large.
 They  examined models with two doublets, a doublet plus a singlet
and also a supersymmetric model where there exist three Higgs
supermultiplets.

Maalampi et al. [7] have recently studied  the two-doublet model
in the same vein  as LQT. They  derived an upper bound of the
neutral Higgs boson mass by a numerical analysis, which gave
them more or less  the same bound as of LQT.
It should be pointed out herewith that they did not consider
a  broad class  of scattering processes to derive constraints
on all of the charged  and neutral Higgs boson masses.

The purpose of the present paper is to reexamine the two-doublet
model to see whether  one can derive an upper bounds for neutral
as well as charged Higgs boson masses in the method of LQT.
  We will answer to this question in the affirmative by taking
into our considerations   sufficiently large number of scattering
processes. Overall allowed regions of three neutral and one charged
Higgs boson masses are explored  and their maximally possible
values are presented (see Eqs. (45)-(48)).
Moreover we will argue that at least one of the Higgs bosons
ought to be lighter considerably than might have  been expected
from the LQT' work (see Eqs. (49) and (50)).

\vskip1cm
\noindent
{\bf 2. Two Higgs Doublet Model}

Let us start by specifying the  $SU(2)_{W}\times U(1)_{Y}$
invariant Higgs potential for two $Y=1$,   $SU(2)_{W}$ doublets,
$\Phi _{1}$ and $\Phi _{2}$.  To avoid the flavor changing neutral
current, we assume the discrete symmetry under $\Phi _{2}
 \rightarrow -\Phi _{2}$ [8].  The most general potential then
consists of  five quartic couplings together with mass terms
$$
\eqalignno{
V(\Phi _{1}, \Phi _{2})&=\sum _{i=1}^{2} \bigl (- \mu _{i}^{2}
\vert \Phi _{i} \vert ^{2}+\lambda _{i}\vert \Phi _{i} \vert ^{4}
\bigr ) +\lambda _{3} \vert \Phi _{1}\vert ^{2} \vert \Phi _{2}
\vert ^{2}\cr
& +\lambda _{4}({\rm Re} \Phi _{1}^{\dag}\Phi _{2})^{2}+
\lambda _{5}({\rm Im} \Phi _{1}^{\dag}\Phi _{2})^{2}.&(1)\cr
}
$$
The spontaneous symmetry breaking is triggered by two vacuum
expectation values, $v_{1}$ and $v_{2}$ of each doublet field
and we write
$$
\eqalign{
\Phi _{i}=\left(\matrix{w_{i}^{+}\cr
             {1 \over \sqrt{2}}(v_{i}+h_{i}+iz_{i})\cr}\right).
}
\eqno{(2)}
$$
In general, we can take $v_{1} > 0$,   and $v_{2} > 0$.

The mass terms in (1) may be diagonalized by rotation
$$
\eqalign{
\left(\matrix{h_{1}\cr
              h_{2}\cr}\right)
=
\left(\matrix{\cos \alpha&-\sin \alpha\cr
              \sin \alpha& \cos \alpha\cr}\right)
\left(\matrix{h\cr
              H\cr}\right),
}\eqno{(3)}
$$
$$
\eqalign{
\left(\matrix{w_{1}\cr
              w_{2}\cr}\right)
=
\left(\matrix{\cos \beta&-\sin \beta\cr
              \sin \beta& \cos \beta\cr}\right)
\left(\matrix{w\cr
              G\cr}\right),
\left(\matrix{z_{1}\cr
              z_{2}\cr}\right)
=
\left(\matrix{\cos \beta&-\sin \beta\cr
              \sin \beta& \cos \beta\cr}\right)
\left(\matrix{z\cr
              \zeta\cr}\right).
}\eqno{(4)}
$$
In fact the mixing angles are determined by
$$
\eqalign{
\tan \alpha ={-\lambda _{1}v_{1}^{2}+\lambda _{2}v_{2}^{2}+
\sqrt{(\lambda_{1}v_{1}^{2}-\lambda_{2}v_{2}^{2})^{2}+
(\lambda_{3}+\lambda_{4})^{2}v_{1}^{2}v_{2}^{2}}
\over
(\lambda_{3}+\lambda_{4})v_{1}v_{2}
}.
}\eqno{(5)}
$$
 $(-\pi/2 \leq \alpha \leq \pi/2)$ and  $\tan \beta =v_{2}/v_{1}$
 ~$(0 <  \beta  <  \pi /2)$.
We note that $w$ and $z$ are massless Nambu-Goldstone bosons
and are absorbed into the longitudinal components of the gauge
bosons.
The masses of the other fields, $h$, $H$, $G$, and $\zeta $ are
expressed in terms of $\lambda _{i}$~~$(i=1,\cdots ,5)$  and
$v_{i}$~~$(i=1,2)$.
Or conversely, the five quartic couplings are given by [6]
$$
\eqalignno{
\lambda _{1}=& {G_{F} \over \sqrt{2}\cos ^{2}\beta }\biggl (
m_{h}^{2}\cos ^{2}\alpha +m_{H}^{2}\sin ^{2}\alpha \biggr),&(6)\cr
\lambda _{2}=& {G_{F} \over \sqrt{2}\sin ^{2}\beta }\biggl (
m_{h}^{2}\sin ^{2}\alpha +m_{H}^{2}\cos ^{2}\alpha \biggr ),&(7)\cr
\lambda _{3}=& {\sqrt{2}G_{F} \sin 2\alpha \over \sin 2\beta}
\biggl (m_{h}^{2}-m_{H}^{2}\biggr )+2\sqrt{2}G_{F}m_{G}^{2},&(8)\cr
\lambda _{4}=& -2\sqrt{2}G_{F}m_{G}^{2},&(9)\cr
\lambda _{5}=& 2\sqrt{2}G_{F}\biggl (m_{\zeta}^{2}-m_{G}^{2}
\biggr ),&(10)\cr
}
$$
where we have set $\sqrt{v_{1}^{2}+v_{2}^{2}}=(\sqrt{2}G_{F}
)^{-1/2}$
and have assumed $\beta \not= 0, \pi /2$ ($v_{1}$,  $v_{2} > 0$).

We will take it for granted hereafter that the quartic couplings
$\lambda _{i}~~~(i=1,\cdots 5)$ are lying in the region
ensuring the positivity of all the masses squared.  The conditions
are summarized by
$$
4\lambda _{1}\lambda _{2} > (\lambda _{3}+\lambda _{4})^{2},
\eqno{(11)}
$$
$$
\lambda _{1}v_{1}^{2}+\lambda _{2}v_{2}^{2} > 0, ~~~\lambda _{5} >
\lambda _{4},~~~0 > \lambda _{4}.
\eqno{(12)}
$$
We will also assume $m_{h}>m_{H}$ without spoiling generality.

\vskip1cm
\noindent
{\bf 3. Eigenvalues of S-Matrix}

The formulae (6) - (10) show that, if the quartic couplings
$\lambda _{i}~~~(i=1,\cdots ,5)$ are comparable to, say, unity,
then the Higgs masses are also on the order of $G_{F}^{-1/2}$.
The upper bounds   of neutral as well as charged Higgs boson
masses are derived by assuming that all these quartic couplings
are within a perturbative region.
As a criterion of the weak quartic couplings we adopt the
perturbative unitarity \`a
 la LQT, thereby constraining the diagonal elements of the
S-matrix.

It has been known that a potential threat to unitarity comes
from longitudinal gauge bosons and Higgs particles.
Scatterings involving the longitiudinal gauge bosons are
replaced in the high energy limit by those of correponding
Nambu-Goldstone bosons thanks to  the equivalence theorem
[1, 9, 10]. Our problem  therefore boils down just to focusing
the Higgs-Goldstone system described by (1).   Furthermore, the
dominant contribution in  the high energy comes only from the
quartic couplings, which we will  consider   henceforth
exclusively.   The S-matrix then becomes independent of
energy, {\it i.e.,}  just a constant.

We are interested in    various two-body scatterings between
fourteen neutral states,
$w^{+}w^{-}$, $w^{+}G^{-}$, $G^{+}w^{-}$, $G^{+}G^{-}$, $zz$,
$z\zeta$, $\zeta \zeta$, $hh$, $hH$, $HH$, $hz$, $Hz$, $\zeta h$,
 and $\zeta H$.
The  interactions in (1) are, however,  extremely involved in
terms of $w$, $G$, $z$, $\zeta$ , $h$, and $H$,  and so will be
the S-matrix, too, not to mention the diagonalization thereof.
It is now important to notice that the S-matrix evaluated in the
mass eigenstate bases can always be transferred into the one in
the original fields $w_{i}$,  $z_{i}$,  and $h_{i}$ ~$(i=1,2)$ by
making a unitary transformation.
The S-matrix in the original field bases is to be calculated
by using the quartic part of the interactions  (1) expressed in
terms of the original fields which looks much simpler of course.
Since all we need to know are the eigenvalues of the S-matrix, it
suffices to deal with the S-matrix in the original field bases.
We are thus able to streamline our calculations just by using the
original fields, $w_{i}$, $z_{i}$, and $h_{i}$~~~$(i=1,2)$ from
the outset.

The use of the original fields are justified by the following
considerations.  In the high energy scattering,  dominant
contributions to the amplitudes come merely from the quartic
coplings as was mentioned before,  {\it i.e.,}  Feynman diagrams
containing triple Higgs couplings are suppressed in energy on the
dimensional account.  The absence of Higgs propagators in our
calculations  amounts to ignoring admittedly the mass
differences  of various Higgs states originating from propagators.
As in the  usual particle mixing among degenerate states (such as
massless neutrinos), we are allowed  to choose freely the most
suitable   field variables.

We  evaluate scattering matrix on the basis of the  states,
$\vert w_{i}^{+}w_{j}^{-}>$,  $\vert z_{i}z_{j}>$,  $\vert
h_{i}h_{j}>$, $\vert h_{i}z_{j}>$ ~~~$(i,j=1,2)$ and/or
combinations thereof.
We will find following orthonormal bases  the most convenient:
$$
\eqalignno{
& \vert A_{i} >={1 \over 2\sqrt {2}}
\vert 2w_{i}^{+}w_{i}^{-}+ z_{i}z_{i}+ h_{i}h_{i}>
,&(13)\cr
& \vert B_{i} >={1 \over 2\sqrt{2}}
\vert 2w_{i}^{+}w_{i}^{-}-  z_{i}z_{i}-  h_{i}h_{i}>
,&(14)\cr
& \vert C_{i} >={1 \over 2}
\vert z_{i}z_{i}-h_{i}h_{i}>,&(15)\cr
& \vert D_{i}>=\vert h_{i}z_{i}>,&(16)\cr
& \vert E_{1}>={1 \over \sqrt {2}i}\vert w_{1}^{+}w_{2}^{-}-
w_{2}^{+}w_{1}^{-}>
,~~~
\vert E_{2}>={1\over \sqrt{2}}\vert h_{1}z_{2}-h_{2}z_{1}>,
&(17)\cr
& \vert F_{\pm} >={1 \over 2}\vert w_{1}^{+}w_{2}^{-}+w_{2}^{+}
w_{1}^{-}\pm z_{1}z_{2}\pm h_{1}h_{2}>,&(18)\cr
& \vert F_{1}>={1 \over \sqrt {2}}\vert z_{1}z_{2}-h_{1}h_{2}>,
{}~~~\vert F_{2}>={1 \over \sqrt{2}}\vert h_{1}z_{2}+h_{2}z_{1}>.
&(19)\cr
}
$$

The tree level high energy scatterings consist dominantly of
S-wave,
 since they are  described by the contact  interactions with
quartic couplings in the above bases.
The transition matrix of   S-wave  amplitudes  turns out to
be of a block-diagonal form  ${\rm  diag} ~(A, B, C, D, E,
f_{+}, f_{-}, f_{1}, f_{2})$
with the following submatrices
$$
\eqalign{
A={1 \over 16\pi}
\left(
\matrix{
6\lambda _{1} &2\lambda _{3}+{1 \over 2}(\lambda _{4}+
\lambda _{5})\cr
2\lambda _{3}+{1 \over 2}(\lambda _{4}+\lambda _{5})&
6\lambda _{2}\cr
}\right),
}
\eqno{(20)}
$$
$$
\eqalign{
B={1 \over 16\pi}
\left(
\matrix{
2\lambda _{1}&{1 \over 2}(\lambda _{4}+\lambda _{5})\cr
{1 \over 2}(\lambda _{4}+\lambda _{5})&2\lambda _{2}\cr
}\right),
}
\eqno{(21)}
$$
$$
\eqalign{
C=D={1 \over 16\pi}
\left(
\matrix{
2\lambda _{1}&{1 \over 2}(\lambda _{4}-\lambda _{5})\cr
{1 \over 2}(\lambda _{4}-\lambda _{5})&2\lambda _{2}\cr
}\right),
}
\eqno{(22)}
$$
$$
\eqalign{
E={1 \over 16\pi}
\left(
\matrix{
\lambda _{3}+{1 \over 2}(3\lambda _{5}-\lambda _{4})&
-\lambda _{5}\cr
-\lambda _{5}&\lambda _{3}+{1\over 2}(3\lambda _{5}-
\lambda _{4})\cr
}\right).
}
\eqno{(23)}
$$
The rows and colums of these $2\times 2$ matrices are spanned
by the states, (13)-(17), respectively.  The diagonal elements
of the amplitudes $f_{\pm}$, $f_{1}$ and $f_{2}$ are those of
(18) and (19) and are given by
$$
\eqalign{
f_{+}={1 \over 16\pi}(\lambda _{3}+{5 \over 2}\lambda _{4}-
{1 \over 2}\lambda _{5}),~~~
f_{-}={1 \over 16\pi}(\lambda _{3}+{1 \over 2}\lambda _{4}-
{1 \over 2}\lambda _{5}),
}\eqno{(24)}
$$
$$
\eqalign{
f_{1}=f_{2}={1 \over 16\pi}(\lambda _{3}+{1 \over 2}\lambda
_{4}+{1\over 2}\lambda _{5}).
}\eqno{(25)}
$$
We are  lucky enough to obtain the fourteen eigenvalues of
the transition matrix  analytically.
The eigenvalues of  (20)-(23) are given respectively as
follows;
$$
\eqalignno{
a_{\pm}=& {1 \over 16\pi}\biggl \{3(\lambda _{1}+\lambda _{2})
\pm \sqrt { 9(\lambda _{1} -\lambda _{2})^{2}+\{2\lambda _{3}+
{1 \over 2}(\lambda _{4}+\lambda _{5})  \}^{2}    }\biggr \},
&(26)\cr
b_{\pm}=& {1 \over 16\pi}\biggl \{ (\lambda _{1}+\lambda _{2})
\pm \sqrt {(\lambda _{1}-\lambda _{2})^{2}+{1 \over 4}(\lambda
_{4} +\lambda _{5})^{2}      }\biggr \},&(27)\cr
c_{\pm}=d_{\pm}=& {1 \over 16\pi}\biggl \{(\lambda _{1}+\lambda
_{2})\pm \sqrt {(\lambda _{1}-\lambda _{2})^{2}+{1 \over 4}(
\lambda _{4}-\lambda _{5})^{2}                  }\biggr \},
&(28)\cr
e_{1}=& {1 \over 16\pi}(\lambda _{3}-{1 \over 2}\lambda _{4}+
{5 \over 2}\lambda _{5}),&(29)\cr
e_{2}=& { 1\over 16\pi}(\lambda _{3}-{1 \over 2}\lambda _{4}+
{1 \over 2}\lambda _{5}). &(30)\cr
}
$$

The reason for the block diagonal form of our transition matrix
is explained in the following way.
 First of all let us recall the discrete symmetry under $\Phi _{2}
 \rightarrow -\Phi _{2}$.  This indicates that the number of suffix
 ``2" is conserved modulo two throughout scattering processes. In
other words, the states $\vert A_{i}>$,  $\vert B_{i}>$,  $\vert
C_{i}>$, and $\vert D_{i}>$   are not able to communicate with
the other states $\vert E_{i}>$,   $\vert F_{\pm}>$, and $\vert
F_{i}>$.

Besides this discrete symmetry the quartic part of Higgs potential
(1) is endowed with the following discrete symmetry
$$
\eqalignno{
C:~~~~~&\Phi _{i} \rightarrow \Phi _{i} ^{\dag},&(31)\cr
Y_{\pi}:~~~~~&\Phi _{i}\rightarrow \exp ({i \over 2}\pi )\Phi _{i}
,&(32)\cr
G:~~~~~&\Phi _{i} \rightarrow \exp ({i \over 2}\pi \sigma _{2})
\Phi _{i}^{\dag}.&(33)
}$$
The various states (13)-(19) are classified according to the
properties under these discrete transformations.   It is
straightforward to see that $\vert A_{i}>$,  $\vert B_{i}>$,
 $\vert C_{i}>$,   $\vert F_{\pm}>$ and $\vert F_{1}>$ are
$C$-even states, while  the others are $C$-odd.    We also
note that $\vert A_{i}>$,  $\vert B_{i}>$,  $\vert E_{i}>$
and $\vert F_{\pm}>$ are $Y_{\pi}$-even and the others are
$Y_{\pi}$-odd.  On the other hand, not all of the states are
 the eigenstates of the $G$-transformation: there are only
two $G$-even states ($\vert A_{i}>$ and $\vert F_{+}>$)  and
two $G$-odd states ($\vert B_{i}>$ and $\vert F_{-}>$).
These discrete symmetries are efficient enough to set up
selection rules and to understand vanishing  elements of
the transition matrix  between those states with different
quantum numbers.

In their analysis of the S-matrix, LQT argued that there was
an  $O(4)$ symmetry in the single-doublet Higgs potential in
the high energy limit. They made a full use of this in close
analogy with isospin symmetry, classifying    states into
$O(4)$ representations.
Although our states (13)-(16) are reminiscent of their
$O(4)$-analysis,   our Higgs potential (1) does not have
such continuous symmetries.

In passing our S-matrix  was once studied partly by Casalubuoni
et al. [6].  They restricted themselves, however,  to a special
 case that $m_{h}$ was much greater than $G_{F}^{-1/2}$ and  the
mixing angle $\alpha$ was negligibly small. Note also that
Maalampi et al. [7] have  studied  scattering processes with
seven elastic channels $ w^{+}w^{-}$, $G^{+}G^{-}$,  $zz$,
$\zeta \zeta$,   $hh$, $hH$, and   $HH$ (diagonal channel
analysis).

\vskip1cm
\noindent
{\bf 4. Analysis of the Tree Unitarity Constraints}

Now that we have the eigenvalues of the  transition matrix,
we are in a position to impose the tree unitarity conditions .
The unitarity is respected if the eigenvalues  are lying within
the so-called Argand circle on the complex plane.
The tree level amplitudes, however, are necessarily real
and therefore outside the  circle.
 If higher order radiative corrections are included, the
amplitudes will eventually settle down within the circle
  and unitarity
 will be restored.  For such perturbative calculations to
be meaningful,  the tree level amplitudes should not be very
far away from the circle.  As a criterion for the   perturbative
recovery of unitarity, LQT argued that the eigenvalues should not
exceed unity,
$$
\vert a_{\pm}\vert,~~~
\vert b_{\pm}\vert,~~~
\vert c_{\pm}\vert,~~~
\vert d_{\pm}\vert,~~~
\vert e_{i}\vert,  ~~~
\vert f_{\pm}\vert,~~~
\vert f_{i}\vert~\leq 1.
\eqno{(34)}
$$
In some literatures [11], the RHS of (34) is replaced by 1/2 , the
radius of the Argand circle.
We, however, use more conservative conditions (34),  LQT's, since
it will be instructive to see how the inclusion of extra Higgs
bosons  will  modify the LQT's result.

For illustration, we take up the first one $
\vert a_{+}\vert \leq 1
$.
If we express $\lambda _{i}$'s in $a_{+}$ in terms of $m_{h}$,
$m_{H}$, $m_{G}$, $m_{\zeta}$, $\alpha$, $\beta$, and $G_{F}$,
with the help of Eqs. (6)-(10), this  inequality provides us
with a relation that must be satisfied by various Higgs
bosons masses together with the mixing angles.
After a little manipulation we find  $\vert a_{+}\vert \leq 1$
  equivalent to the following two conditions
$$
\eqalignno{
& {9 \over 9-5\sin ^{2}2\alpha} \biggl (X-{8\pi \over 3}\biggr )
^{2}-\biggl (Y-Y_{0}\biggr )^{2}\geq R ^{2},&(35)\cr
& {8\pi \over 3}\sin ^{2}2\beta
\geq
X-Y\cos 2\alpha \cos 2\beta ,
&(36)\cr
}
$$
where our notations are
$$
\eqalignno{
X=& {1 \over \sqrt {2}}G_{F}(m_{h}^{2}+m_{H}^{2}),&(37)\cr
Y=& {1 \over \sqrt{2}}G_{F}(m_{h}^{2}-m_{H}^{2}),&(38)\cr
Y_{0}=& {1 \over 9-5\sin ^{2}2\alpha }\biggl \{24\pi
\cos 2\alpha \cos 2\beta -\sqrt{2}G_{F}(m_{\zeta }^{2}+
2m_{G}^{2})\sin 2\alpha
\sin 2\beta \biggr \},&(39)\cr
R =&{ 1\over 9-5\sin ^{2}2\alpha }\biggl \{16\pi \sin
2\alpha \cos 2
\beta   +{3 \over \sqrt{2}}G_{F}(m_{\zeta}^{2}+2m_{G}^{2})
\sin 2\beta
\cos 2\alpha \biggr \}.&(40)\cr
}$$
The second inequality (36) excludes the right half of the hyperbola
(35)  on  the   $(X, Y)$ plane    and the allowed region is the
shaded one in Fig. 1, where    $Y\geq 0$ ~($m_{h} > m_{H}$) and
$X-Y=\sqrt{2}G_{F}m_{H}^{2}\geq 0$ are  taken into account.

Fig. 1 shows clearly that the allowed region of $m_{H}$ and
$m_{h}$ is bounded for fixed values of mixing algles,
$m_{\zeta }$ and $m_{G}$.
We can read off  bounds for $m_{H}$ and $m_{h}$
 from the shaded region
$$
\eqalignno{
X-Y& =\sqrt{2}G_{F}m_{H}^{2}\leq X_{P},&(41)\cr
X+Y& =\sqrt{2}G_{F}m_{h}^{2}\leq 2X_{Q},&(42)\cr
}$$
where $X_{P}$ and $X_{Q}$ are the $X$-coordinates of the points
$P$ and $Q$,  respectively in Fig. 1:
$$
\eqalignno{
X_{P} = &{8\pi \over 3}-{1 \over 3}\sqrt{(9-5 \sin ^{2}2\alpha )
(Y_{0}^{2}+R^{2})},&(43)\cr
X_{Q} = &{1 \over 5\sin ^{2} 2\alpha}\biggl [24\pi -(9-5\sin
^{2}2\alpha )Y_{0}\cr
&-\sqrt{(9-5\sin ^{2}2\alpha )\{5R^{2}\sin ^{2}2\alpha +(8\pi
-3Y_{0})^{2}\}
}
\biggr ]
.&(44)\cr
}
$$

We have similarly analyzed other eigenvalues, and found after
 all that (41) and (42) are the most stringent conditions upon
$m_{H}$ and $m_{h}$.
Since the points $P$ and $Q$ are both on the left half of
the hyperbola,  $X_{P}\leq 8\pi/3$ and $X_{Q}\leq 8\pi/3$
should hold in general.  We thus have upper bounds for neutral
Higgs boson masses
$$
\eqalignno{
& m_{H}\leq \biggl ({4\pi\sqrt{2} \over 3G_{F}}\biggr )^{1/2}
={1 \over \sqrt{2}}M_{LQT},&(45)\cr
& m_{h}\leq \biggl ({8\pi\sqrt{2} \over 3G_{F}}\biggr )^{1/2}
=M_{LQT},&(46)\cr
}
$$
whatsoever the values of $\alpha$, $\beta$, $m_{\zeta}$ and
$m_{G}$.

The relations (41) and (42) contain more information than (45)
and (46).
 To elucidate this,
 let us have a closer look at  (41).  In the $(G_{F}m_{H}^{2},
G_{F}m_{G}^{2}, G_{F}m_{\zeta}^{2})$-space, The inequality  (41)
 tells us that  the inner region surrounded by the solid curves
in Fig. 2 are  allowed exclusively.
  This means that the charged Higgs boson masses $m_{G}$ and
$m_{\zeta}$ are also bounded. In fact Fig. 2 shows apparently
$$
\eqalignno{
m_{G}& \leq \biggl ({4\pi \sqrt{2} \over G_{F}}
\biggr )^{1/2}=\sqrt{{3 \over 2}}M_{LQT},&(47)\cr
m_{\zeta}& \leq \biggl ({8\pi \sqrt{2} \over G_{F}}\biggr )^{1/2}
=\sqrt{3}M_{LQT}
.&(48)
}
$$
The most interesting bound is derived by  a geometrical
inspection of Fig. 2.    We obtain the bound on the lightest
Higgs boson mass among $m_{H}$, $m_{G}$,  and $m_{\zeta }$,
\vfill\eject
$$
\eqalignno{
M({\rm the~ lightest~ Higgs~ boson~ mass}) & \leq
\sqrt{{\sin ^{2}2\beta \over 4-\sin ^{2}2\beta
}
}M_{LQT}\cr
& =\sqrt{{x^{2} \over 1+x^{2}+x^{4}}}M_{LQT},&(49)\cr
}
$$
where $x=v_{2}/v_{1}=\tan \beta$.  This is our main result.
The upper limit in (49) is realized at point K on the surface
in Fig. 2 where the vector $OK$ is parallel to (1, 1, 1).

\vskip1cm
\noindent
{\bf 5. Summary }

In the present paper we have investigated the consequences of
applying the tree unitarity conditions to   the two Higgs
doublet model.
  We have seen that, by considering a wide class of scattering
processes,
not only neutral but also charged Higgs boson masses are
bounded from above. Our results are listed in Eqs. (45)-(48).

The most important results of our calculations is  that at least
 one of the Higgs bosons should be much lighter than has been
anticipated from   LQT's work [1].  The maximum value of the RHS
of (49) is
reached when $x=1$~($\sin ^{2}2\beta =1$).
It is therefore concluded that the lightest Higgs boson mass has
to satisfy
$$
M({\rm the~lightest~ Higgs ~boson ~mass})
\leq {1 \over \sqrt{3}}M_{LQT}=580 ~{\rm GeV}/c^{2},
\eqno{(50)}
$$
that is, the bound is  $1/\sqrt{3}$ of the one derived by  LQT
for a single doublet Higgs case.
It is very likely that inclusion of more Higgs doublets would
give us more tight upper bound.

A comment to be made herewith  is that, if we would  use $\vert
a_{+}\vert \leq 1/2$ as a criterion of perturbative recovery of
unitarity  [11],  all the bounds of Higgs masses  that we have
obtained would be reduced by a factor $1/\sqrt{2}$.  In
particular, the upper bound of the lightest Higgs boson
mass (50) would be replaced by $M_{LQT}/\sqrt{6}=$ 410
GeV/$c^{2}$.

Our final remark is that, so far, we have assumed $v_{1}$,
$v_{2}$ $\not=$ 0.  If one of these vacuum expectation values
e.g. $v_{2}$ happens to vanish,  then the neutral Higgs boson
$h_{2}$  becomes massless at the tree level.  This is a pseudo-
Nambu-Goldstone boson because there does not exist any
particular symmetry prohibitting   $h_{2}$ from  acquiring
a mass on the loop level.  Then it is expected that $v_{2}$
becomes non-zero by the higher loop corrections as far as
$\Phi _{2}$ is coupled with $\Phi _{1}$.
\vskip2cm
\noindent
{\bf ACKNOWLEDGEMENTS}

Our sincere thanks should go to Takeshi Kurimoto for
invaluable discussions.  One of us (E.T.) is supported in
part by Grant in Aid for  Scientific Research,
from the Ministry of Education, Science and Culture
(No. 02640230).

\vskip2cm
\noindent
{\bf REFERENCES}
\item{[1]}
B.W. Lee, C. Quigg and H.B. Thacker, Phys. Rev. Lett.
{\bf 38} (1977) 883; Phys. Rev. {\bf D16} (1977) 1519.
\item{[2]}
D.A. Dicus and V.S. Mathur, Phys. Rev. {\bf D7}  (1973) 3111.
\item{[3]}
S. Weinberg, Phys. Rev. Lett. {\bf 37} (1976) 657.
\item{[4]}
R.D. Peccei and H.R. Quinn, Phys. Rev. Lett. {\bf 38} (1977) 1440.
\item{[5]}
A.J. Buras, P. Krawczyk, M.E. Lautenbacher and C. Salazar,
Nucl. Phys. {\bf B337} (1990) 284;
V. Barger, J.L. Hewett and R.J.N. Phillips, Phys. Rev.
{\bf D41} (1990) 3421;
J.F. Gunion and B. Gradkowski, Phys. Lett. {\bf B245} (1991)
 591;
J.F. Gunion, H.E. Haber,  G. Kane ans S. Dawson,
``{\it The Higgs Hunter's Guide }"  (Addison-Wesley Pub.
Co. 1990)
\item{[6]}
R. Casalbuoni, D. Dominici, F. Feruglio and R. Gatto, Nucl.
Phys. {\bf B299} (1988) 117;  Phys. Lett. {\bf B200} (1988) 495;
R. Casalbuoni, D. Dominici,  R. Gatto and C. Giunti, Phys.
Lett. {\bf B178} (1986) 235;
\item{[7]}
J. Maalampi, J. Sirkka and I. Vilja, Phys. Lett. {\bf B265}
(1991) 371.
\item{[8]}
S.L. Glashow and S. Weinberg, Phys. Rev. {\bf D15} (1977) 1958.
\item{[9]}
J.M. Cornwall, D.N. Levin and G. Tiktopoulos, Phys. Rev.
 Lett. {\bf 30} (1973) 1268; Phys. Rev. {\bf D10} (1974) 1145.
\item{[10]}
M.S. Chanowitz and M.K. Gaillard, Nucl.  Phys. {\bf B261} (1985)
379;
Y.P. Yao and C.P. Yuan, Phys. Rev. {\bf D38} (1988) 2237;
M. Veltman, Phys. Rev. {\bf D41} (1990) 2294;
J. Bagger and C. Schmit, Phys. Rev. {\bf D41} (1990) 264;
 H.J.   He, Y-P. Kuang and X. Li,  Phys. Rev. Lett.
  {\bf 69}  (1992) 2619;
K. Aoki, in  {\it  Proceedings  of the Meeting on
Physics at TeV Energy Scale}
(KEK Report 89-20,  Nov. 1987) p. 20.
\item{[11]}
W. Marciano, G. Valencia, and S. Willenbrock,  Phys. Rev.
{\bf D40}  (1989) 1725;
M. L\"uscher and P. Weisz,   Phys. Lett. {\bf B212}
 (1989) 472.

%\vskip2cm
\vfill\eject
\noindent
{\bf FIGURE CAPTIONS}
\item{Fig. 1}
The allowed region of Higgs boson masses given by (35)
and (36). Our variables are $X=G_{F}(m_{h}^{2}+m_{H}^{2})
/\sqrt{2}$ and $Y=G_{F}(m_{h}^{2}-m_{H}^{2})/\sqrt{2}$.
\item{Fig. 2}
Schematic view of the allowed region of Higgs boson
masses $m_{H}$, $m_{\zeta }$ and $m_{G}$   given by Eq.
 (41) for given values of mixing angles.  The inner region
surrounded by the solid curves is allowed.
The vector $OK$ is in the direction of (1, 1, 1).

\end